\documentclass[twocolumn,runinaddress,showpacs,prb]{revtex4}
\usepackage{graphicx,bm}

\begin{document}

\title{Spin resolved Hall effect driven by spin-orbit coupling}
\author{Jian Li$^{1}$, Liangbin Hu$^{2}$, and Shun-Qing Shen$^{1}$}
\affiliation{$^{1}$Department of Physics, The University of Hong Kong, Pukfulam Road,
Hong Kong, China\\
$^{2}$Department of Physics, South China Normal University, Guangzhou
510631, China}
\date{January 18, 2005}

\begin{abstract}
Spin and electric Hall currents are calculated numerically in a
two-dimensional mesoscopic system with Rashba and Dresselhaus spin-orbit
coupling by means of the Landauer-B\"{u}ttiker formalism. It is found that
both electric and spin Hall currents circulate when two spin-orbit couplings
coexist, while the electric Hall conductance vanishes if either one is
absent. The electric and spin Hall conductances are suppressed in strong
disorder, but survive in weak disorder. Physically it can be understood that
the spinomotive transverse "force" generated by spin-orbit coupling is
responsible for the formation of the spin Hall current and the lack of
transverse reflection symmetry is the origin of the electric Hall current.
\end{abstract}

\maketitle

When a metallic sample is subjected to a perpendicular external magnetic
field, the Lorentz force acting on the charge carriers gives rise to a
transverse voltage between two edges of the sample, this is well known as
the ordinary Hall effect. Actually the Hall effect family has numbers of
important members. The anomalous Hall effect may occur even in the absence
of an external magnetic field in a ferromagnetic metal with spin-orbit
coupling.\cite{Chien79, Karplus54, Smit55, Berger70, Onoda02JPSJ,
Jungwirth02PRL, Onoda03prl, Fang03} In the past few years it has been
recognized that the spin-orbit coupling may provide an efficient way to
manipulate a spin resolved current in metals and semiconductors.\cite%
{Dyakonov71, Hirsch99prl, Zhang00prl, Hu03prb, Murakami03Science,
Sinova04prl, Shen04prl} In a two dimensional electron gas (2DEG) lacking
bulk and structure inversion symmetries, the effective Hamiltonian is given
by
\begin{equation}
H=\frac{p^{2}}{2m^{\ast }}+\lambda (\sigma ^{x}p_{y}-%
\sigma ^{y}p_{x})+\beta (\sigma ^{x}p_{x}-\sigma ^{y}p_{y})
\label{ham}
\end{equation}%
where the second term is the Rashba spin-orbit coupling and the third one is
the Dresselhaus spin-orbit coupling. $\sigma ^{\mu }$ ($\mu =x,y,z$) are the
Pauli matrices and the coupling parameters $\lambda $ and $\beta $ have the
dimension of velocity. Using the Heisenberg equation of motion the second
derivative of the position operator $\mathbf{r}$ gives
\begin{equation}
m^{\ast }\frac{\partial ^{2}\mathbf{r}}{\partial t^{2}}=+\frac{2m^{\ast
}(\lambda ^{2}-\beta ^{2})\sigma ^{z}}{\hbar }\mathbf{p}\times \hat{z}.
\label{force}
\end{equation}%
Compared with the Lorentz force\ brought by the magnetic field
upon a charged particle, the spin-orbit coupling produces a
spinomotive transverse "force" on a moving electron. It has no
classical counterpart as the coefficient is divided by $\hbar $,
but it reflects the tendency of spin asymmetric scattering of a
moving electron subject to the spin-orbit coupling. When charge
carriers are driven by an external electric field, this "force"
tends to form a transverse spin current.

In this paper we calculate the spin and electric Hall conductances in a 2DEG
mesoscopic system with Rashba and Dresselhaus coupling by using the
Landauer-B\"{u}ttiker formula and the Green's function technique. It is
found that both electric and spin Hall currents circulate while these two
types of spin-orbit coupling coexist, but the electric Hall current
disappears when either one is absent. The spin and electric Hall
conductances are suppressed in strong disorder, but survive in weak
disorder. The numerical results are in good agreement with the symmetry
analysis of the system.
\begin{figure}[tbp]
\begin{center}
\includegraphics[width=0.35\textwidth] {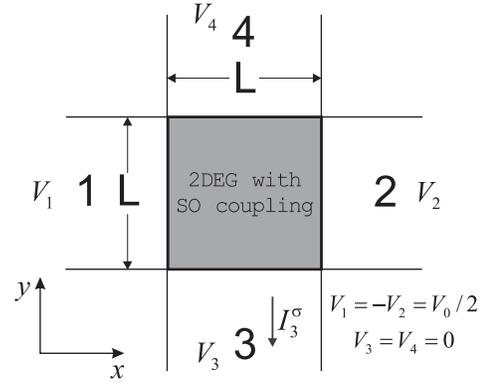}
\end{center}
\caption{Cross-shape device with four semi-infinite metallic leads. The
spin-orbit coupling is supposed to exist in the shadowed area only, and the
effect of the semi-infinite leads is treated exactly through self-energy
terms.}
\label{setup}
\end{figure}

We consider a cross-shape device with four semi-infinite metallic leads as
shown in Fig. 1. The scattering region (shadowed part in Fig. 1) is
described by the effective Hamiltonian in Eq.(\ref{ham}), and when it is
treated as an $L\times L$ lattice with the tight-binding approximation, the
model Hamiltonian reads,
\begin{eqnarray}
H &=&-t\sum\limits_{ij,\sigma =\uparrow ,\downarrow }c_{i,\sigma
}^{\dag }c_{j,\sigma }+t_{so}^{R}\sum\limits_{i}\left(
J_{i,y}^{x}-J_{i,x}^{y}\right)
\nonumber \\
&&+t_{so}^{D}\sum\limits_{i}\left( J_{i,x}^{x}-J_{i,y}^{y}\right)
\label{model}
\end{eqnarray}%
where $t_{so}^{R}$ and $t_{so}^{D}$ are the dimensionless parameters for
Rashba and Dresselhaus coupling strength in the unit of $t$, respectively,
and the local spin current operator $J_{i,\alpha }^{\mu }$ is defined as
\cite{Shen97pla}
\begin{equation}
J_{i,\alpha }^{\mu }=-it\sum\limits_{\sigma ,\sigma ^{\prime
}}\left( c_{i,\sigma }^{\dag }\sigma _{\sigma \sigma ^{\prime
}}^{\mu }c_{i+\alpha ,\sigma ^{\prime }}-h.c.\right) .
\end{equation}%
where $\alpha $ stands for the unit vector along axes of the lattice and $%
\mu $ stands for the direction of spin polarization.\cite{note-unit}

The calculation of electric and spin currents is based on the Landauer-B\"{u}%
ttiker formalism.\cite{Buttiker86prl, Datta95} Assume
$T_{q,p}^{\nu ,\mu }$ to be the spin-resolved transmission
probability of electrons transmitted from spin channel $\mu $ of
lead $p$ to spin channel $\nu $ of lead $q$, and $V_{p}$ to be the
electric voltage in lead $p$, then respectively the outgoing
electric current and spin current polarized along $\mu $ direction
in lead $p$ are
\begin{eqnarray}
I_{p}^{c} &=&\frac{e^{2}}{h}\sum_{q,\mu ,\nu }(T_{p,q}^{\mu ,\nu
}V_{q}-T_{q,p}^{\nu ,\mu }V_{p}), \\
I_{p}^{\mu } &=&-\frac{e}{4\pi }\sum_{q,\nu }\left[ (T_{p,q}^{\mu ,\nu
}-T_{p,q}^{-\mu ,\nu })V_{q}-(T_{q,p}^{\nu ,\mu }-T_{q,p}^{\nu ,-\mu })V_{p}%
\right] .  \label{spincur}
\end{eqnarray}%
The transmission probability coefficients can be calculated by
using the Green's function technique, $T_{q,p}^{\nu ,\mu
}=$Tr$\left[ \Gamma _{q}^{\nu }G^{R}\Gamma _{p}^{\mu }G^{A}\right]
.$ And the retarded and advanced Green functions are given by
$G^{R,A}(E)=1/(E-H_{c}-\sum_{p=1}^{4}\Pi _{p}^{R,A})$, where $E$
is the electron energy and $H_{c}$ is the model Hamiltonian in the
shadowed region in Fig.1. The retarded and advanced self energy
terms introduced by the semi-infinite lead $p$,\ $(\Pi
_{p}^{R})_{p_{i}\sigma ,p_{j}\sigma ^{\prime }}=-t\sum_{m}\chi
_{m}(p_{i})e^{ik_{m}a}\chi _{m}(p_{j})\delta _{\sigma \sigma
^{\prime }}$ and $\Pi _{p}^{A}=(\Pi _{p}^{R})^{\dag }$ where $\chi
_{m}(p_{i})$ is the transverse mode wave function at site $p_{i}$
in lead $p$ connected to the scattering region. It should be noted
that in Eq.(\ref{spincur}) $\mu $ may stand for an arbitrary
direction of spin polarization, and this is done by incorporating
a transformation in the definition of $\Gamma _{p}$, that is,
$\Gamma _{p}^{\mu }(p_{i},\sigma ,p_{j},\sigma ^{\prime
})=2tR_{\mu \sigma }R_{\sigma ^{\prime }\mu }^{-1}\sum_{m}\chi
_{m}(p_{i})\sin (k_{m}a)\chi
_{m}(p_{j})$ and $R$ is the rotation matrix to transform the eigenstates of $%
\sigma ^{z}$ to those of $\hat{r}\cdot \bm{\sigma}$ ($\hat{r}$ is a unit
vector).\cite{note} The Landauer-B\"{u}ttiker formalism has been applied
extensively to study the spin transport in mesoscopic systems numerically.%
\cite{Pareek04prl, Hankiewicz04prb, Sheng05prl, Nikolic04xxx}
\begin{figure}[tbp]
\centering \includegraphics[width=0.5\textwidth]{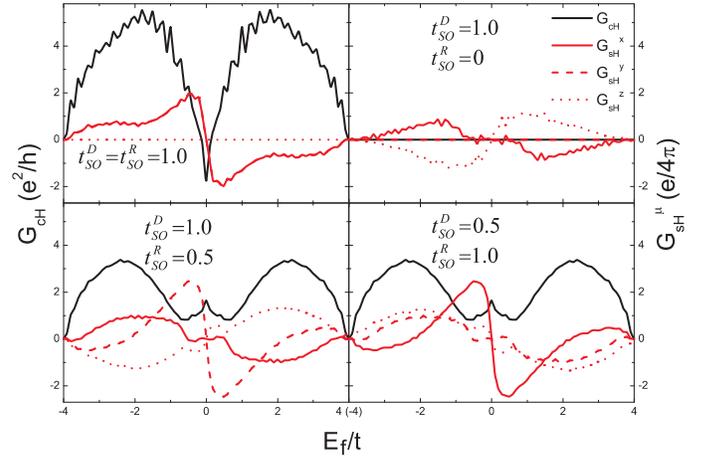}
\caption{Electric and spin Hall conductances as functions of the electron
Fermi energy for different ratios of Rashba and Dresselhaus coupling
strength. $L=40$ for all the results in this figure.}
\label{spin-charge}
\end{figure}

In this paper we consider an initial electric current driven through leads 1
and 2, $I_{1}^{c}=-I_{2}^{c},$ by setting the bias voltage $%
V_{1}=-V_{2}=V_{0}/2$ and $V_{3}=V_{4}=0.$ The currents in leads 3 and 4 are
perpendicular to the current through leads 1 and 2. Thus the electric and
spin Hall conductances are defined as
\begin{eqnarray}
G_{cH} &=&I_{3}^{c}/(V_{1}-V_{2}); \\
G_{sH}^{\mu } &=&I_{3}^{\mu }/(V_{1}-V_{2}),
\end{eqnarray}%
respectively, where the spin current has three components, $\mu =x,y,z.$
Electric and spin Hall conductances are evaluated as functions of the Fermi
energy for different ratios of Rashba and Dresselhaus coupling strength in
Fig. \ref{spin-charge}. Generally speaking, the electric Hall conductance is
symmetric about the Fermi energy $E_{f}$ while the spin Hall conductance is
antisymmetric such that the spin Hall conductance vanishes at the band
center, $E_{f}=0$. This is consistent with the symmetry analysis for the
tight binding Hamiltonian.\cite{Sheng05prl} In the case of the pure Rashba
or Dresselhaus coupling, the electric Hall conductance disappears, but the
spin Hall conductance still exists. In the two cases of $t_{SO}^{R}=1$ and $%
t_{SO}^{D}=1/2$ and of $t_{SO}^{R}=1/2$ and $t_{SO}^{D}=1,$ the
electric Hall conductances are equal. However the spin Hall
conductances $G_{sH}^{z}$ differ by a minus sign, with
$G_{sH}^{x}$ and $G_{sH}^{y}$ swapped, and the former is
consistent with Shen\ and Sinitsyn et al's works for free 2DEG
systems.\cite{Shen04prb, Sinitsyn04prb} A special case is at the
symmetric point of $t_{SO}^{D}=t_{SO}^{R}$. The spin Hall
conductance $G_{sH}^{z}=0$, while $G_{sH}^{x}$ and $G_{sH}^{y}$
are equal and non-zero, which means the current is polarized
within the x-y plane. In this case the operator $\sigma
^{x}+\sigma ^{y}$ commutes with the total Hamiltonian, and
actually there is no spin flip in the scattering
region.\cite{Schliemann03prl} On the other hand the longitudinal
electric and spin conductances are also non-zero. The longitudinal
conductances are about one order larger than the Hall conductances
in magnitude, i.e., $I_{3}^{c}/I_{2}^{c}\sim 10.$ And the electric
conductance is also symmetric with respect to the Fermi energy,
just like the electric Hall conductance, while the longitudinal
spin current
is antisymmetric. According to the symmetry properties of such a system,\cite%
{Shen04prb} under the transformation: $\sigma ^{x}\rightarrow \sigma
^{y},\sigma ^{y}\rightarrow \sigma ^{x},$ and $\sigma ^{z}\rightarrow
-\sigma ^{z},$ $t_{SO}^{D}\rightarrow t_{SO}^{R}$ and $t_{SO}^{R}\rightarrow
t_{SO}^{D}$, correspondingly $G_{sH}^{x}\rightarrow G_{sH}^{y},$ $%
G_{sH}^{y}\rightarrow G_{sH}^{x},$ and $G_{sH}^{z}$ $\rightarrow -G_{sH}^{z}$
while the electric Hall conductance remains unchanged. At the symmetric
point, $t_{SO}^{R}=t_{SO}^{D},$ it yields that $G_{sH}^{x}$ $=G_{sH}^{y}$
and $G_{sH}^{z}$ $=0.$ Our numerical results obviously agree with this
symmetry analysis.

\begin{figure}[tbp]
\begin{center}
\includegraphics[width=0.45\textwidth]{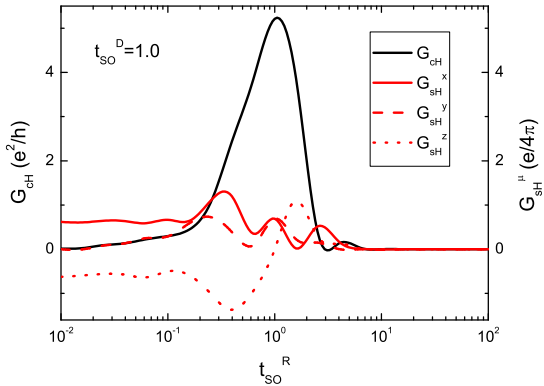}
\end{center}
\caption{Electric and spin Hall conductances as functions of the Rashba
coupling strength with a fixed Dresselhaus coupling strength $t_{so}^{D}=1.0$
at $E_f=-2.0t$ and $L=40$. Similar results are obtained for the Hall
conductances as functions of the Dresselhaus coupling strength with a fixed
Rashba coupling strength, which are consistent with symmetry analyses. }
\label{strength}
\end{figure}

The Hall conductances as functions of the Rashba coupling strength
are also evaluated, with specific Dresselhaus coupling strength
$t_{so}^{D}=1.0$ and electron Fermi energy $E_{f}=-2.0t$ as shown
in Fig. \ref{strength}. It indicates clearly that the electric
Hall conductance increases with
increasing the Rashba couple strength, and reaches its maximal value at $%
t_{so}^{R}=t_{so}^{D}$. Then it turns to decrease when $%
t_{so}^{R}>t_{so}^{D},$ and approaches to zero for a large Rashba coupling
strength. The figure shows that $G_{sH}^{z}=0$ and $G_{sH}^{x}=G_{sH}^{y}$
at $t_{so}^{R}=t_{so}^{D}$ as expected by the symmetry analysis. For a large
spin-orbit coupling both electric and spin Hall conductance approaches to
zero because the spin-orbit coupling in the scattering region forms a large
potential barrier and the incident electrons will be completely reflected.
Unlike bulk systems\cite{Shen04prb, Sinitsyn04prb} where the spin Hall
conductance in the clean limit has a universal value $\pm e/8\pi $ and the
sign is given by the relative ratio of two coupling strength in Eq.(\ref{ham}%
), Fig. \ref{strength} shows that the value of spin Hall conductance varies
with the relative ratio as well as the sign, but the change of sign is
compatible with the bulk systems case. And this result is also compatible
with the previous numerical work in the case of pure Rashba coupling.\cite%
{Sheng05prl, Nikolic04xxx, Nomura05prb}

\begin{figure}[tbp]
\begin{center}
\includegraphics[width=0.5\textwidth]{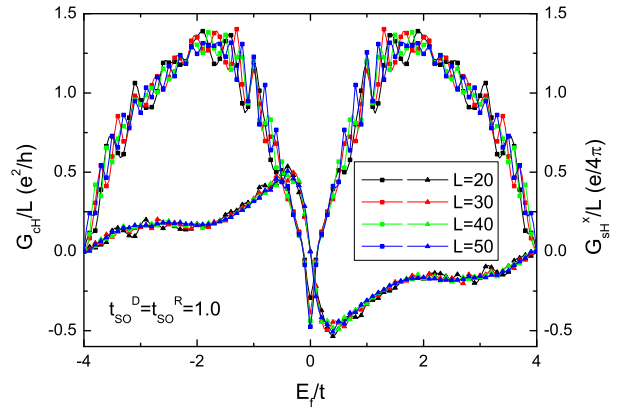}
\end{center}
\caption{Electric and spin Hall conductances divided by sample size $L$ as
functions of the electron Fermi energy. Here the Rashaba and Dresselhaus
coupling strength are equal and the results for $L=20$, $30$, $40$, and $50$
are shown simultaneously. It implies a size effect that both Hall
conductances are proportional to $L$ in this calculation.}
\label{size}
\end{figure}

To see the finite size effect we calculate the electric and spin Hall
conductances for $L=20$, $30$, $40$, and $50$. $G_{cH}/L$ and $G_{sH}/L$ as
functions of $E_{f}$ are plotted in Fig. \ref{size}. We notice that these
curves for different sizes fit a single one very well. Thus we conclude that
both electric and spin Hall conductance are proportional to the size $L$ of
the sample. In other words, in our calculation the electric and spin Hall
currents are determined by both the number of the incident channels and that
of the outgoing channels. Thus the Hall currents induced by a specified
longitudinal electric field are not proportional to the size $L$ linearly,
but to $L\times L$.

\begin{figure}[tbp]
\begin{center}
\includegraphics[width=0.45\textwidth]{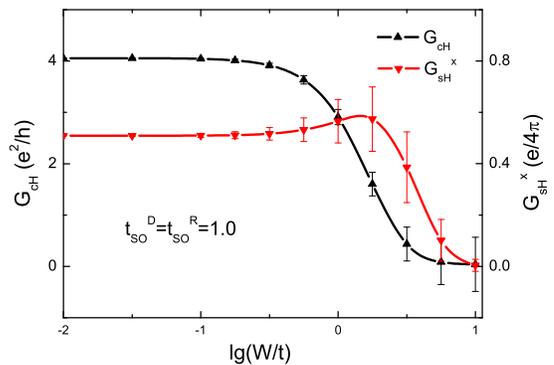}
\end{center}
\caption{Electric and spin Hall conductances as functions of logarithm of
the disorder strength $W/t$. Results are obtained with equal Rashba and
Dresselhaus coupling strength at $E_f=-2.0t$ and $L=30$. Standard deviations
in the calculation are shown through the error bars.}
\label{disorder}
\end{figure}

The disorder effect is an interesting issue in the spin Hall effect in 2DEG.
It is still greatly controversial whether the spin Hall effect may survive
when the impurity scattering is taken into account.\cite{Inoue04prb,
Mishchenko04prl, Rashba04prb} We consider the disorder effect by including
the disorder potential term $V_{disorder}=\sum_{i,\sigma =\uparrow
,\downarrow }\epsilon _{i}c_{i,\sigma }^{\dag }c_{i,\sigma }$ in Eq.(\ref%
{model}) where $\epsilon _{i}$ are randomly distributed between $[-W/2,+W/2]$%
. Selectively the electric and spin Hall conductances, $G_{cH}$ and $%
G_{sH}^{x},$ for two couplings with equal strength are plotted in Fig. \ref%
{disorder}. $G_{sH}^{z}$ is exactly equal to zero according to the
symmetry. It shows that both electric and spin Hall conductances
can survive in weak disorder, but be suppressed in strong
disorder. We also did calculation for several other cases, and
obtained similar results. The case of pure Dresselhaus coupling is
in agreement with Sheng et al's work for pure Rashba
coupling.\cite{Sheng05prl}

Physically the spin Hall conductance can be well understood from the
spinomotive transverse "force" caused by the spin-orbit coupling in Eq.(\ref%
{force}). The electric field drives electrons moving along the field such
that the electrons with spin-up or -down experience opposite transverse
"force" and thus a non-zero spin current is induced perpendicular to the
field. The relative ratio of the two coupling strength determines the
direction of the spin Hall current as the spinomotive force changes its sign
around $\lambda =\beta $ and vanishes at the point. All calculated results
are consistent with this. However, the spinomotive force is not a direct
origin of the non-zero $G_{cH}$, since $G_{cH}$ arises only when two
couplings are present simultaneously. From the symmetry properties of the
system we notice that the Hamiltonian with pure Rashba coupling is invariant
under the transformation: $k_{x}\rightarrow k_{x},$ $k_{y}\rightarrow -k_{y}$
and $\sigma ^{x}\rightarrow -\sigma ^{x},$ $\sigma ^{y}\rightarrow \sigma
^{y},$ $\sigma ^{z}\rightarrow -\sigma ^{z}$. Similarly the Hamiltonian with
pure Dresselhaus coupling is invariant under the transformation: $%
k_{x}\rightarrow k_{x},$ $k_{y}\rightarrow -k_{y}$ and $\sigma
^{x}\rightarrow \sigma ^{x},$ $\sigma ^{y}\rightarrow -\sigma
^{y},$ $\sigma ^{z}\rightarrow -\sigma ^{z}$. This is why the
electric Hall current vanishes in these two cases, while the spin
Hall current circulates because there is no symmetry constraint on
it as both $k_{y}$ and $\sigma ^{z}$ change their signs under such
transformation. On the other hand, the Hamiltonian with both
Rashba and Dresselhaus couplings does not possess the reflection
symmetry of $k_{x}\rightarrow k_{x},$ $k_{y}\rightarrow -k_{y}$.
Therefore the coexistence of both couplings breaks the reflection
symmetry of the system, which makes the electric current not
parallel to the electric field such that it gives rise to a
nonvanishing Hall conductance $G_{cH}$. This unconventional Hall
conductance may be related to some discussions in terms of the
anomalous Hall effect due to parity anomaly and additional band
crossing\cite{Onoda02JPSJ}. Moreover, since the diagonal spin
conductance is non-zero in this case\cite{Sinitsyn04prb}, the
diagonal spin current along leads 1 and 2 might generate a charge
Hall current via the reciprocal spin Hall effect.\cite{Zhang05xxx}

In conclusion, we studied the electric Hall conductance as well as
the spin Hall conductance for a finite-size system with four
leads. Both electric and spin Hall conductances are non-zero when
both Rashba and Dresselhaus coupling are present, thus the current
is actually spin polarized. Unlike the anomalous Hall effect, the
present electric Hall current is driven by the spin-orbit
coupling, not by the exchange coupling with the magnetic
impurities.\cite{Culcer03prb} This effect also differs from the
one resulted from a spin polarized current via the Rashba
coupling.\cite{Bulgakov99prl} Though the incident current is not
spin polarized, the Hall current is polarized in our case.

The authors would like to thank L. Sheng and D. N. Sheng for helpful
discussions. This work was supported by the Research Grant Council of Hong
Kong (SQS), and by the National Science Foundation of China under Grant No.:
10474022 (LBH).


\begin{thebibliography}{99}
\bibitem{note-unit} The unit of the spin current operator is $at$ ($a$ is
the lattice space), and here we take $a=1$.
\bibitem{note} We take $+\mu $ as up ($\uparrow $) and $-\mu $ as down ($
\downarrow $) as the indices of the rotation matrix $R.$
\end{thebibliography}
\end{document}